\documentclass{article} 
\usepackage{spconf}
\usepackage{graphicx,psfrag}
\usepackage{url,hyperref}
\usepackage[usenames]{color}
\usepackage[cmex10]{amsmath}
\usepackage{cite}
\usepackage{amsfonts,amssymb,latexsym,cite,color,multirow,rotating}
\usepackage{algorithm,algorithmic}
\usepackage{bbm}
\usepackage[normalem]{ulem} 

 \newcommand{\figsize}{0.95\columnwidth}
 \newcommand{\putTable}[3]{\begin{table}[t!]
  			    \centering
		            #3
			    \vspace{-2mm}
     			    \caption{#2}
     			    \label{tab:#1}
			    \vspace{-2mm}
			  \end{table} }
 \newcommand{\putFrag}[4]{\begin{figure}[t!]
                            \begin{center}
                            #4
			    \includegraphics[width=#3]{figures/#1.eps}
			    \vspace{-3mm}
            		    \caption{#2}
           		    \label{fig:#1}
			    \vspace{-3mm}
                            \end{center}
                          \end{figure} }

 \renewcommand{\hat}{\widehat}
 \renewcommand{\tilde}{\widetilde}
 \newcommand{\defn}{\triangleq}

 \newcommand{\hvec}[1]{\ensuremath{\boldsymbol{\Hat{#1}}}}
    
 \renewcommand{\vec}[1]{\ensuremath{\boldsymbol{#1}}}
 \newcommand{\mat}[1]{\ensuremath{\begin{bmatrix}#1\end{bmatrix}}}

 \newcommand{\mc}[1]{\ensuremath{\mathcal{#1}}}

 \newcommand{\Real}{{\mathbb{R}}}

 \newcommand{\tran}{^\textsf{T}}
 \newcommand{\herm}{^\textsf{H}}

 \newcommand{\ind}{1\hspace{-3.5pt}\mbox{1}}

 \DeclareMathOperator{\E}{E}

 \DeclareMathOperator*{\argmax}{arg\, max}
 \DeclareMathOperator*{\argmin}{arg\, min}

 \renewcommand{\eqref}[1]{(\ref{eq:#1})}
 
 \newcommand{\Figref}[1]{Figure~\ref{fig:#1}}
 
 \newcommand{\figref}[1]{Fig.~\ref{fig:#1}}
 
 \newcommand{\tabref}[1]{Table~\ref{tab:#1}}
 \newcommand{\secref}[1]{Sec.~\ref{sec:#1}}

 \newcounter{comment}[section]
 
 \newcounter{texthead}[section]

 \newcommand{\MAP}{\textsf{MAP}}
 \newcommand{\MMSE}{\textsf{MMSE}}
 \newcommand{\SNR}{\textsf{SNR}}
 \newcommand{\NMSE}{\textsf{NMSE}}
 \newcommand{\NSNR}{\textsf{NSNR}}

 \newcommand{\U}{\textsf{u}}
 
 \newcommand{\vQ}{\textsf{\textbf{\textit{q}}}}

 \newcommand{\vX}{\textsf{\textbf{\textit{x}}}}
 
 \newcommand{\vY}{\textsf{\textbf{\textit{y}}}}
 
 \newcommand{\vZ}{\textsf{\textbf{\textit{z}}}}

\begin{document}
\ninept
\setlength{\arraycolsep}{0.4mm}
\title{Generalized Approximate Message Passing for Cosparse Analysis Compressive Sensing}
\twoauthors{Mark Borgerding and Philip Schniter\thanks{This work was supported by NSF grants CCF-1018368, CCF-1218754, and an allocation of computing time from the Ohio Supercomputer Center.}}
{Dept. ECE, The Ohio State University\\
 Columbus, OH 43210}
{Sundeep Rangan}
{Dept. of ECE, NYU Polytechnic Institute\\
 Brooklyn, NY 11201.}
		
 \date{\today}
 \maketitle

\begin{abstract}
In cosparse analysis compressive sensing (CS), one seeks to estimate a non-sparse signal vector from noisy sub-Nyquist linear measurements by exploiting the knowledge that a given linear transform of the signal is cosparse, i.e., has sufficiently many zeros.
We propose a novel approach to cosparse analysis CS based on the generalized approximate message passing (GAMP) algorithm.
Unlike other AMP-based approaches to this problem, ours works with a wide range of analysis operators and regularizers.
In addition, we propose a novel $\ell_0$-like soft-thresholder based on MMSE denoising for a spike-and-slab distribution with an infinite-variance slab.
Numerical demonstrations on synthetic and practical datasets demonstrate advantages over existing AMP-based, greedy, and reweighted-$\ell_1$ approaches.
\end{abstract}

\begin{keywords}
Approximate message passing, belief propagation, compressed sensing.
\end{keywords}

\section{Introduction} \label{sec:intro}

We consider the problem of recovering a signal $\vec{x}\in\Real^N$ (e.g., an $N$-pixel image) from the possibly noisy linear measurements
\begin{align}
\vec{y} 
&= \vec{\Phi x}+\vec{w} \in\Real^M ,
\label{eq:y}
\end{align}
where $\vec{\Phi}$ represents a known linear measurement operator $\vec{w}$ represents noise, and $M\ll N$.
We focus on the \emph{analysis compressive sensing} (CS) problem \cite{Elad:IP:07,Nam:ACHA:13} where, for a given analysis operator $\vec{\Omega}$,
\begin{align}
\vec{u}
&\defn \vec{\Omega}\vec{x} \in \Real^D \label{eq:u}
\end{align}
is assumed to be \emph{cosparse} (i.e., contain sufficiently many zero-valued coefficients).
This differs from the \emph{synthesis CS} problem, where $\vec{x}$ is assumed to be \emph{sparse} (i.e., contain sufficiently few non-zero coefficients). 
Although the two problems become interchangeable when $\vec{\Omega}$ is invertible, we are mainly interested in non-invertible $\vec{\Omega}$, as in the ``overcomplete'' case where $D>N$. 
We note that, although we assume real-valued quantities throughout, the proposed methods can be directly extended to the complex-valued case, which we demonstrate using numerical experiments.

The analysis CS problem is typically formulated as a regularized loss-minimization problem of the form 
\begin{align}
\hvec{x}_{\textsf{rlm}} 
&= \argmin_{\vec{x}} \tfrac{1}{2}\|\vec{y}-\vec{\Phi}\vec{x}\|_2^2 + h(\vec{\Omega}\vec{x}) ,
\label{eq:ana_lam}
\end{align}
with separable regularizer $h(\vec{u})=\sum_{d=1}^D h_d(u_d)$.
One of the most famous instances of $h(\vec{u})$ is that of total-variation (TV) regularization \cite{Rudin:PhyD:92}, where $h(\vec{u})=\lambda\|\vec{u}\|_1$ and $\vec{\Omega}$ computes variation across neighboring pixels.
In the anisotropic case, this variation is measured by finite difference operators, e.g., $\vec{\Omega}=[\vec{D}_h\herm,\vec{D}_v\herm]\herm$, where $\vec{D}_h$ computes horizontal differences and $\vec{D}_v$ computes vertical differences.
Of course, $\ell_1$ regularization can be used with generic $\vec{\Omega}$, with the desirable property that it always renders \eqref{ana_lam} convex.
The resulting problem, sometimes referred to as the \emph{generalized LASSO} (GrLASSO) \cite{Tibshirani:AS:11}, is amenable to a wide range of efficient optimization techniques like Douglas-Rachford splitting \cite{Combettes:JSTSP:07} 
and NESTA \cite{Becker:JIS:11}. 

Despite the elegance of the $\ell_1$ norm, several studies have shown improvements from the use of $\ell_0$-like norms for $h(\vec{u})$, especially for highly overcomplete $\vec{\Omega}$ (i.e., $D\gg N$).
For example, the use of iteratively reweighted $\ell_1$ \cite{Candes:JFA:08} has demonstrated significant improvements over $\ell_1$ regularization in the context of analysis CS \cite{Carrillo:SPL:13,Chartrand:ASIL:13}.
Likewise, greedy approaches to locate the zero-valued elements in $\vec{u}$ have also demonstrated significant improvements over $\ell_1$. 
Examples include greedy analysis pursuit (GAP), analysis iterative hard thresholding (AIHT), analysis hard thresholding pursuit (AHTP), analysis CoSaMP (ACoSaMP), and analysis subspace pursuit (ASP) \cite{Nam:ACHA:13,Giryes:12}.

In this paper, we propose a Bayesian approach to analysis CS that leverages recent advances in approximate message passing (AMP) algorithms \cite{Donoho:PNAS:09,Donoho:ITW:10a}, and in particular the generalized AMP (GAMP) algorithm from \cite{Rangan:ISIT:11}.

While other AMP-based approaches have been recently proposed for the special case where $\vec{\Omega}$ is a 1D finite difference operator, i.e., the TV-AMP from \cite{Donoho:TIT:13b} and the ssAMP from \cite{Kang:14}, our approach works with a \emph{generic} analysis operator $\vec{\Omega}$ and a much broader range of signal priors and likelihoods.
Furthermore, our approach facilitates both MAP and (approximate) MMSE estimation of $\vec{x}$ in a computationally efficient manner.
We also note that a different Bayesian approach to cosparse analysis CS, based on multivariate Gauss-mixture priors, was recently presented in \cite{Turek:DSP:14}. 
The MAP and MMSE estimation methods proposed in \cite{Turek:DSP:14}, which employ greedy pursuit and Gibbs sampling, respectively, have computational complexities that scale as $O(D^2 N)$ and $O(D^2 N^3)$ (assuming $M\leq D$).
In contrast, ours scales like $O(DN)$ for generic $\vec{\Omega}$, or $O(N\log N)$ when $\vec{\Phi}$ and $\vec{\Omega}$ have fast implementations, which is often the case in imaging applications.

\section{Generalized AMP for Analysis CS} \label{sec:approach}

\subsection{The proposed Bayesian model} \label{sec:model}

Our approach is Bayesian in that it treats the true signal $\vec{x}$ as a realization of a random vector $\vX\in\Real^N$ with prior pdf $p_{\vX}(\vec{x})$ and likelihood function $p_{\vY|\vQ}(\vec{y}|\vec{\Phi x})$, where $\vec{y}$ are the observed noisy measurements and $\vQ \defn \vec{\Phi}\vX$ are akin to hidden noiseless measurements.
(For clarity, we write random quantities using san-serif fonts and deterministic ones using serif fonts.)
Furthermore, we assume that the prior and likelihood have the forms 
\begin{align}
p_{\vY|\vQ}(\vec{y}|\vec{\Phi x})
&\propto \prod_{m=1}^M \exp(-l_m([\vec{\Phi x}]_m))
\label{eq:like}  \\
p_{\vX}(\vec{x})
&\propto \prod_{d=1}^D \exp(-h_d([\vec{\Omega}\vec{x}]_d)) \prod_{n=1}^N \exp(-g_n(x_n))
\label{eq:prior}
\end{align} 
with scalar functions $l_m(\cdot)$, $h_d(\cdot)$, and $g_n(\cdot)$.
Note that each measurement value $y_m$ is coded into the corresponding function $l_m(\cdot)$.
We discuss the design of these functions in the sequel.

Given the form of \eqref{like} and \eqref{prior}, the MAP estimate $\hvec{x}_\MAP \defn \argmax_{\vec{x}}p_{\vX|\vY}(\vec{x}|\vec{y})$ 
can be written (using Bayes rule) as 
\begin{align} 
\hvec{x}_\MAP
&= \arg\min_{\vec{x}} \big\{
l(\vec{\Phi}\vec{x})
+ h(\vec{\Omega}\vec{x})
+ g(\vec{x})  \big\}
\label{eq:MAP}
\end{align} 
with separable loss function $l(\vec{q})\!=\!\sum_{m=1}^M l_m(q_m)$ and
separable regularizers $g(\vec{x})\!=\!\sum_{n=1}^N g_n(x_n)$ and
$h(\vec{u})\!=\!\sum_{d=1}^D h_d(u_d)$.
Note that, with trivial $g(\vec{x})\!=\!0$ and quadratic loss $l(\vec{q})\!=\!\frac{1}{2}\|\vec{q}-\vec{y}\|_2^2$, the MAP estimation problem \eqref{MAP} reduces to the regularized loss minimization problem \eqref{ana_lam}.
But clearly \eqref{MAP} is more general.

As for the MMSE estimate $\hvec{x}_\MMSE \defn \int \vec{x}\, p_{\vX|\vY}(\vec{x}|\vec{y}) d\vec{x}$, exact evaluation requires the computation of a high dimensional integral, which is intractable for most problem sizes of interest.
In the sequel, we present a computationally efficient approach to MMSE estimation that is based on loopy belief propagation and, in particular, the GAMP algorithm from \cite{Rangan:ISIT:11}.


\subsection{Background on GAMP} \label{sec:gamp}

The GAMP algorithm \cite{Rangan:ISIT:11} aims to estimate the signal $\vec{x}$ from the corrupted observations $\vec{y}$, where $\vec{x}$ is assumed to be a realization of random vector $\vX\!\in\!\Real^N$ with known prior $p_{\vX}(\vec{x})$ and likelihood function $p_{\vY|\vZ}(\vec{y}|\vec{Ax})$. 
Here, the prior and likelihood are assumed to be separable in the sense that
\begin{equation}
p_{\vY|\vZ}(\vec{y}|\vec{z})
\propto \prod_{i=1}^I \exp(-f_i(z_i))
,\quad
\label{eq:pYZ}
p_{\vX}(\vec{x})
\propto \prod_{n=1}^N \exp(-g_n(x_n)) ,
\end{equation} 
where $\vZ\!\defn\!\vec{A}\vX\!\in\!\Real^{I}$ can be interpreted as hidden transform outputs.
The MAP version of GAMP aims to compute $\hvec{x}_\MAP=\arg\max_{\vec{x}} p_{\vX|\vY}(\vec{x}|\vec{y})$, i.e., solve the optimization problem
\begin{align}
\hvec{x}_\MAP
&= \arg\min_{\vec{x}} \sum_{i=1}^I f_i([\vec{Ax}]_i) + \sum_{n=1}^N g_n(x_n) ,
\label{eq:gamp}
\end{align}
while the MMSE version of GAMP aims to compute the MMSE estimate $\hvec{x}_\MMSE \defn \int \vec{x}\, p_{\vX|\vY}(\vec{x}|\vec{y}) d\vec{x}$, in both cases by iterating simple, scalar optimizations.
MAP-GAMP can be considered as the extension of the AMP algorithm \cite{Donoho:PNAS:09} from the quadratic loss $f(\vec{z})=\|\vec{y}-\vec{z}\|_2^2$ to generic separable losses of the form $f(\vec{z})=\sum_{i=1}^I f_i(z_i)$.
Likewise, MMSE-GAMP can be considered as a similar extension of the Bayesian-AMP algorithm \cite{Donoho:ITW:10a} from additive white Gaussian noise (AWGN) $p_{\vY|\vZ}(\vec{y}|\vec{z})$ to generic $p_{\vY|\vZ}(\vec{y}|\vec{z})$ of the form in \eqref{pYZ}.

In the large-system limit (i.e., $I,N\rightarrow \infty$ with $I/N$ converging to a positive constant) under i.i.d sub-Gaussian $\vec{A}$, GAMP is characterized by a state evolution whose fixed points, when unique, are Bayes optimal \cite{Javanmard:II:13,Bayati:ISIT:12}.
For generic $\vec{A}$, it has been shown \cite{Rangan:ISIT:13} that MAP-GAMP's fixed points coincide with the critical points of the cost function \eqref{gamp} and that MMSE-GAMP’s fixed points coincide with those of a certain variational cost that was connected to the Bethe free entropy in \cite{Krzakala:ISIT:14}.
However, with general $\vec{A}$ (e.g., non-zero-mean $\vec{A}$ \cite{Caltagirone:ISIT:14} or ill-conditioned $\vec{A}$ \cite{Rangan:ISIT:14}) GAMP may not converge to its fixed points, i.e., it may diverge.
In an attempt to prevent divergence with generic $\vec{A}$, damped \cite{Schniter:ALL:12,Rangan:ISIT:14}, adaptively damped \cite{Vila:ICASSP:15}, and sequential \cite{Krzakala:ISIT:14} versions of GAMP have been proposed.

\putTable{gamp}{The damped GAMP algorithm.  In (R4) and (R11), $F'_i$ and $G'_n$ denote the derivatives of $F_i$ and $G_n$ w.r.t their first arguments.}
{\footnotesize
\[\arraycolsep=1pt
\begin{array}{|l@{}rcl@{}r|}\hline 
 \multicolumn{4}{|l}{\textsf{definitions for MMSE-GAMP:}}&\\
 &F_i(\hat{p},\nu^p)
  &\defn& \displaystyle \frac{\int z \exp(-f_i(z)) \mc{N}(z;\hat{p},\nu^p) dz}{\int \exp(-f_i(z)) \mc{N}(z;\hat{p},\nu^p) dz} &\text{\scriptsize(D1)}\\
 &G_n(\hat{r},\nu^r)
  &\defn& \displaystyle \frac{\int x \exp(-g_n(x)) \mc{N}(x;\hat{r},\nu^r) dx}{\int \exp(-g_n(x)) \mc{N}(x;\hat{r},\nu^r) dx} &\text{\scriptsize(D2)}\\[3mm]
 \multicolumn{4}{|l}{\textsf{definitions for MAP-GAMP:}}&\\
 &F_i(\hat{p},\nu^p)
  &\defn& \argmin_{z} f_i(z) + \frac{1}{2\nu^p}|z-\hat{p}|^2 &\text{\scriptsize(D3)}\\
 &G_n(\hat{r},\nu^r)
  &\defn& \argmin_{x} g_n(x) + \frac{1}{2\nu^r}|x-\hat{r}|^2 &\text{\scriptsize(D4)}\\[1mm]
  \hline
  \multicolumn{2}{|l}{\textsf{inputs:}}&&&\\
  &\multicolumn{4}{l|}{\quad \forall i,n\!: F_i, G_n, \hat{x}_n(1), \nu^x_n(1), a_{in}, T_{\max}\geq 1, \epsilon\geq 0, \beta_0\in(0,1]}\\[1mm]
  \multicolumn{2}{|l}{\textsf{initialize:}}&&&\\
  &\multicolumn{4}{l|}{\quad \forall i\!: \hat{s}_i(0)=0, ~ t=1}\\[1mm]
  \multicolumn{4}{|l}{\textsf{for $t =1,\dots, T_{\max}$,}}&\\
  &\multicolumn{3}{l}{\hspace{5mm} \textsf{if} \hspace{1.5 mm}
        t=1,~\textsf{then}~\beta=1,~\textsf{else}~\beta=\beta_0}&\text{\scriptsize(R1)}\\
  &\forall i\!: 
   \nu^p_i(t)
   &=& \beta \textstyle \sum_{n=1}^{N} |a_{in}|^2 \nu^x_{n}(t) + (1\!-\!\beta\big) \nu^p_i(t\!-\!1) &\text{\scriptsize(R2)} \\
  &\forall i\!: 
   \hat{p}_i(t)
   &=& \sum_{n=1}^{N} a_{in} \hat{x}_{n}(t) - \nu^p_i(t) \,\hat{s}_i(t\!-\!1) &\text{\scriptsize(R3)}\\
  &\forall i\!: 
   \nu^z_i(t)
   &=& \nu^p_i(t)\, F'_i(\hat{p}_i(t),\nu^p_i(t)) &\text{\scriptsize(R4)}\\
  &\forall i\!: 
   \hat{z}_i(t)
   &=& F_i(\hat{p}_i(t),\nu^p_i(t)) &\text{\scriptsize(R5)}\\
  &\forall i\!: 
   \nu^s_i(t)
   &=& \displaystyle \beta\left(1\!-\!\frac{\nu^z_i(t)}{\nu^p_i(t)}\right)\frac{1}{\nu^p_i(t)} \!+\! \big(1\!-\!\beta\big)\nu^s_i(t\!-\!1) &\text{\scriptsize(R6)}\\
  &\forall i\!: 
   \hat{s}_i(t)
   &=& \displaystyle \beta\frac{\hat{z}_i(t)-\hat{p}_i(t)}{\nu^p_i(t)} \!+\! \big(1\!-\!\beta\big)\hat{s}_i(t\!-\!1)&\text{\scriptsize(R7)}\\
  &\forall n\!: 
   \tilde{x}_n(t)
   &=&  \beta\hat{x}_n(t) + \big(1\!-\!\beta\big)\tilde{x}_n(t\!-\!1) &\text{\scriptsize(R8)}\\
  &\forall n\!: 
   \nu^r_n(t)
   &=& \displaystyle \beta\left(\frac{1}{\sum_{i=1}^{I} |a_{in}|^2 \nu^s_i(t)}\right) \!+\! \big(1\!-\!\beta\big)\nu^r_n(t\!-\!1) 
        &\text{\scriptsize(R9)}\\
  &\forall n\!: 
   \hat{r}_n(t)
   &=& \textstyle \tilde{x}_n(t)+ \nu^r_n(t) \sum_{i=1}^{I} \!a_{in}^*
        \hat{s}_i(t)  &\text{\scriptsize(R10)}\\
  &\forall n\!: 
   \nu^x_n(t\!+\!\!1)
   &=& \nu^r_n(t)\, G'_n(\hat{r}_n(t),\nu^r_n(t)) &\text{\scriptsize(R11)}\\
  &\forall n\!: 
   \hat{x}_{n}(t\!+\!\!1)
   &=& G_n(\hat{r}_n(t),\nu^r_n(t)) &\text{\scriptsize(R12)}\\
  &\multicolumn{3}{l}{\hspace{5mm} \textsf{if} \hspace{1.5 mm}
        \|\hvec{x}(t)-\hvec{x}(t\!+\!1)\|/\|\hvec{x}(t\!+\!1)\|
        < \epsilon, \textsf{then~stop}} &\text{\scriptsize(R13)}\\
  \multicolumn{2}{|l}{\textsf{end}}&&&\\[1mm]
  \multicolumn{5}{|l|}{\textsf{outputs:~}
        \forall n\!:
        \hat{x}_n(t\!+\!1) 
        }\\[1mm]
  \hline
\end{array}
\]
\vspace{-2mm}
}

A damped version of the GAMP algorithm is summarized in \tabref{gamp}. 
There, smaller values of the damping parameter $\beta_0$ make GAMP more robust to difficult $\vec{A}$ at the expense of convergence speed, and $\beta_0=1$ recovers the original GAMP algorithm from \cite{Rangan:ISIT:11}.
Note that the only difference between MAP-GAMP and MMSE-GAMP is the definition of the scalar denoisers in (D1)-(D4). 
Denoisers of the type in (D3)-(D4) are often referred to ``proximal operators'' in the optimization literature. 
In fact, as noted in \cite{Rangan:ISIT:13} and \cite{Rangan:ISIT:14}, max-sum GAMP is closely related to primal-dual algorithms from convex optimization, such as the classical Arrow-Hurwicz and recent Chambolle-Pock and primal-dual hybrid gradient algorithms \cite{Esser:JIS:10,Chambolle:JMIV:11,He:JIS:12}. 
The primary difference between MAP-GAMP and those algorithms is that the primal and dual stepsizes (i.e., $\nu^r_n(t)$ and $1/\nu^p_i(t)$ in \tabref{gamp}) are adapted, rather than fixed or scheduled.

\subsection{GAMP Enables Analysis CS} \label{sec:grampa}

If we configure GAMP's transform $\vec{A}$ and 
loss function $f(\cdot)$ as
\begin{equation} 
\vec{A} = \mat{\vec{\Phi}\\\vec{\Omega}}, 
~~ 
f_i(\cdot) = 
\begin{cases}
l_i(\cdot) & i\in\{1,\dots,M\} \\
h_{i-M}(\cdot) & i\in\{M\!+\!1,\dots,M\!+\!D\} \\
\end{cases}
\label{eq:augment}
\end{equation} 
where 
$\vec{\Phi}$ and $\vec{\Omega}$ 
are the measurement and analysis operators from \secref{model},
and 
$l_i(\cdot)$ and $h_{d}(\cdot)$ 
are the loss and regularization functions from \secref{model},
then MAP-GAMP's optimization problem \eqref{gamp} coincides with the MAP optimization \eqref{MAP},
which (as discussed earlier) is a generalization of the analysis-CS problem \eqref{ana_lam}.
Likewise, MMSE-GAMP will return an approximation of the MMSE estimate $\hvec{x}_\MMSE$ under the statistical model \eqref{like}-\eqref{prior}.

In the sequel, we refer to GAMP under \eqref{augment} (with suitable choices of $f_q$, $g_n$, and $h_d$) as ``Generalized AMP for Analysis CS,'' or GrAMPA.
Despite the simplicity of this idea and its importance to, e.g., image recovery, it has (to our knowledge) not been proposed before, outside of our preprint \cite{Borgerding:13}.

\subsection{Choice of loss and regularization} \label{sec:design}

One of the strengths of GrAMPA is the freedom to choose the loss function $l_i(\cdot)$ and the regularizations $g_n(\cdot)$ and $h_d(\cdot)$.

The quadratic loss $l_m(q)=|y_m-q|^2$, as used in \eqref{ana_lam}, is appropriate for many applications.  
GrAMPA, however, also supports non-quadratic losses, as needed for 1-bit compressed sensing \cite{Kamilov:TSP:12}, phase retrieval \cite{Schniter:ALL:12}, and Poisson-based photon-limited imaging \cite{Snyder:JASA:93}.

The pixel regularization $g_n(\cdot)$ could be used to enforce 
known positivity in $x_n$ (via $g_n(x)=-\ln \ind_{x\geq 0}$), 
real-valuedness in $x_n$ despite complex-valued measurements (via $g_n(x)=-\ln \ind_{x\in\Real} ~\forall n$), 
or 
zero-valuedness in $x_n$ (via $g_n(x)=-\ln \ind_{x=0} ~\forall n$).
Here, we use $\ind_{A}\in\{0,1\}$ to denote the indicator of the event $A$.

As for the analysis regularization $h_d(\cdot)$, the use of $h_d(u)=\lambda |u|$ with MAP-GrAMPA would allow it to tackle the GrLASSO and anisotropic TV problems defined in \secref{intro}.
With MMSE-GrAMPA, a first instinct might be to use the Bernoulli-Gaussian (BG) prior commonly used for synthesis CS, i.e.,
$h_d(u)= -\ln\big( (1-\beta)\delta(u) + \beta\mc{N}(u;0,\sigma^2)\big)$,
where $\delta(\cdot)$ is the Dirac delta pdf and the parameters
$\beta$ and $\sigma^2$ control sparsity and variance, respectively.
But the need to tune two parameters is inconvenient, and bias effects from the use of finite $\sigma^2$ can degrade performance, especially when $D\gg N$.

Thus, for the MMSE case, we propose a \emph{sparse non-informative parameter estimator} (SNIPE) that can be understood as the MMSE denoiser for a ``spike-and-slab'' prior with an infinite-variance slab.
In particular, SNIPE computes the MMSE estimate of random variable $\U_d$ from a $\mc{N}(0,\nu^q_d)$-corrupted observation $\hat{q}_d$ under the prior
\begin{align}
p_{\U_d\!}(u) 
&= \beta_d\, p_0(u/\sigma)/\sigma + (1-\beta_d) \delta(u) ,
\label{eq:BU}
\end{align}
in the limiting case that $\sigma\rightarrow\infty$. 
Here, $\beta_d\in(0,1]$ is the prior probability that $\U_d\neq 0$ 
and the ``slab'' pdf $p_0(u)$ is continuous, finite, and non-zero at $u=0$, but otherwise \emph{arbitrary}. 
Note that, for fixed $\sigma$ and $\beta_d$, the MMSE estimator can be stated as
\begin{eqnarray}
\lefteqn{
F_d(\hat{q}_d;\nu_d^q)
\defn \E\{\U_d|\hat{q}_d;\nu^q_d\}
= \frac{\int u \,p_{\U_d\!}(u) \mc{N}(u;\hat{q}_d,\nu^q_d) du}
{\int p_{\U_d\!}(u) \mc{N}(u;\hat{q}_d,\nu^q_d) du} \label{eq:denoise}
}\nonumber\\
&=& \frac{ \int u p_0(u/\sigma)\mc{N}(u;\hat{q}_d,\nu^q_d) du}
{\int p_0(u/\sigma)\mc{N}(u;\hat{q}_d,\nu^q_d) du + \sigma\frac{1-\beta_d}{\beta_d}\mc{N}(0;\hat{q}_d,\nu^q_d) }. \qquad
\label{eq:uhat_a}
\end{eqnarray}
Since, with any fixed sparsity $\beta_d<1$, the estimator \eqref{uhat_a} trivializes to $F_d(\hat{q}_d;\nu_d^q)=0~\forall \hat{q}_d$ as $\sigma\rightarrow\infty$,
we scale the sparsity with $\sigma$ as
$
\beta_d
=\sigma/\big(\sigma+p_0(0)\sqrt{2\pi\nu_d^q} \exp( \omega)\big)
$
for a tunable parameter $\omega\in\Real$, in which case it can be shown that
\begin{align}
F_d(\hat{q}_d;\nu^q_d,\omega)
&\stackrel{\sigma\rightarrow\infty}{=} 
\frac{\hat{q}_d}{1+\exp( \omega - \tfrac{1}{2}|\hat{q}_d|^2/\nu^q_d) } .
\label{eq:snipe1}
\end{align}

\section{Numerical results} \label{sec:sims}

We now provide numerical results that compare GrAMPA
with SNIPE denoising to several existing algorithms for cosparse analysis CS. 
In all cases, recovery performance was quantified using $\NSNR \!\defn\! \|\vec{x}\|^2/\|\hvec{x}-\vec{x}\|^2$.
Each algorithm was given perfect knowledge of relevant statistical parameters (e.g., noise variance) or in cases were an algorithmic parameter needed to be tuned (e.g., GrLASSO $\lambda$ or SNIPE $\omega$), the $\NSNR$-maximizing value was used.

\subsection{Comparison to ssAMP and TV-AMP}

We first replicate an experiment from the ssAMP paper \cite{Kang:14}. 
Using the demonstration code for \cite{Kang:14}, we generated signal realizations $\vec{x}\!\in\!\Real^N$ that yield BG 1D-finite-difference sequences $\vec{\Omega x}$ with sparsity rate $0.05$.
Then we attempted to recover those signals from AWGN-corrupted observations 
$\vec{y}=\vec{\Phi x}+\vec{w}\in\Real^M$,
at an $\SNR\defn \|\vec{\Phi x}\|_2^2/\|\vec{w}\|_2^2$ of $60$~dB,
generated with i.i.d Gaussian measurement matrices $\vec{\Phi}$.

\Figref{mse_ssamp_cmp4_60_1} shows median $\NMSE$ versus sampling ratio $M/N$ for ssAMP, TV-AMP, and GrAMPA, over $100$ problem realizations.
There we see GrAMPA uniformly outperforming ssAMP, which uniformly outperforms TV-AMP.
We attribute the performance differences to choice of regularization: GrAMPA's SNIPE regularization is closer to $\ell_0$ than ssAMP's BG-based regularization, which is closer to $\ell_0$ than TV-AMP's $\ell_1$ regularization.
We note the performance of GrAMPA in \figref{mse_ssamp_cmp4_60_1} is much better than that reported in \cite{Kang:14} due to the misconfiguration of GrAMPA in \cite{Kang:14}.

\putFrag{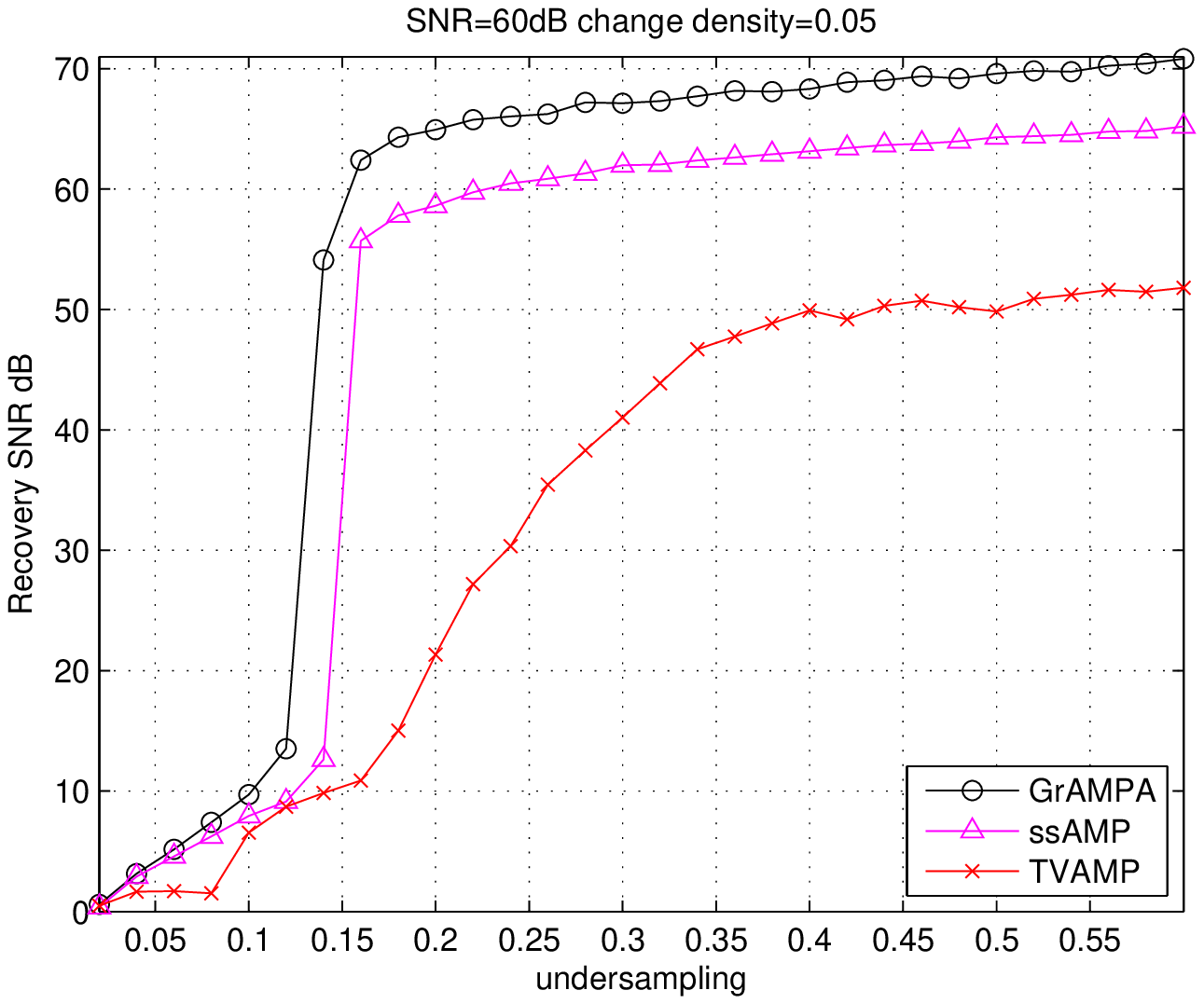}
{Recovery of $0.05$-sparse Bernoulli-Gaussian finite-difference signals from AWGN-corrupted measurements at $\SNR=60$~dB.}
{\figsize}
{\psfrag{undersampling}[t][][0.8]{\sf sampling ratio $M/N$}
 \psfrag{Recovery SNR dB}[b][][0.8]{\sf median recovery NSNR [dB]}
 \psfrag{SNR=60dB change density=0.05}[b][][0.8]{}
 }
{}

\subsection{Comparison to GAP: Synthetic cosparse recovery}

We now compare GrAMPA with SNIPE denoising to Greedy Analysis Pursuit (GAP) \cite{Nam:ACHA:13} using an experiment
from \cite{Nam:ACHA:13} that constructed $\vec{\Omega}\tran\in\Real^{N\times D}$ as a random, almost-uniform, almost-tight frame and $\vec{x}$ as an exactly $L$-cosparse vector.
The objective was then to recover $\vec{x}$ from noiseless measurements $\vec{y}=\vec{\Phi x}$ using analysis operator $\vec{\Omega}$ and i.i.d Gaussian $\vec{\Phi}$.
For this experiment, we used $N=200$.

\begin{figure}[t]
 \begin{center}
   \psfrag{rho,GAP and GrAMPA sigma = 1.0}[b][][0.9]{$\rho=(N\!-\!L)/M$} 
   \psfrag{rho,GAP and GrAMPA sigma = 1.2}[b][][0.9]{} 
   \psfrag{rho,GAP and GrAMPA sigma = 2.0}[b][][0.9]{} 
   \psfrag{GAP and GrAMPA sigma = 1.0}[B][][0.9]{$D/N=1.0$}
   \psfrag{GAP and GrAMPA sigma = 1.2}[B][][0.9]{$D/N=1.2$}
   \psfrag{GAP and GrAMPA sigma = 2.0}[B][][0.9]{$D/N=2.0$}
   \psfrag{delta}[t][][0.9]{$\delta=M/N$} 
   \newcommand{\wid}{0.49\columnwidth}
   \newcommand{\wwid}{0.48\columnwidth}
   \begin{tabular}{@{}c@{}c@{}c@{}}
   \includegraphics[width=\wwid]{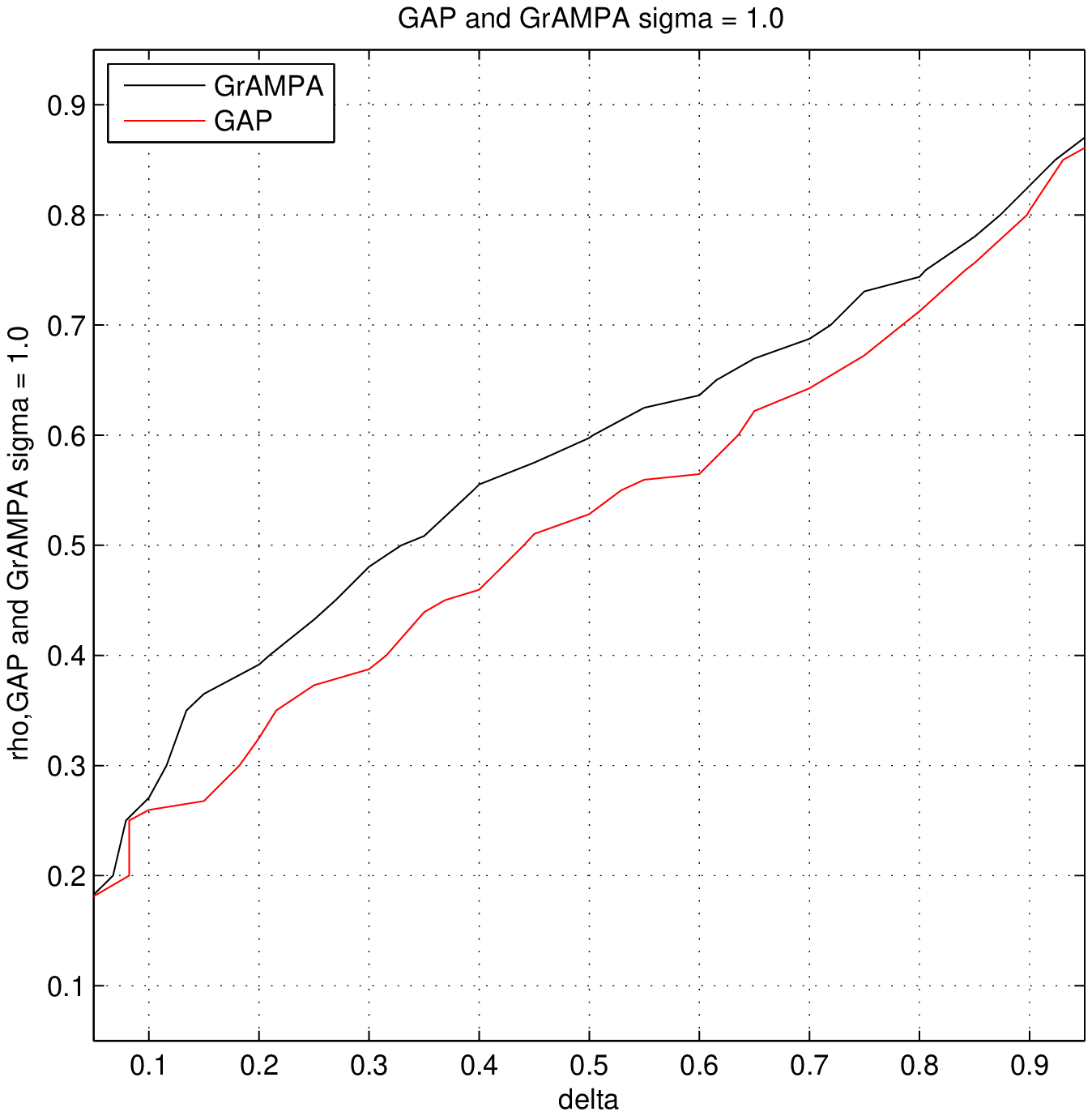}&
   \includegraphics[width=\wwid]{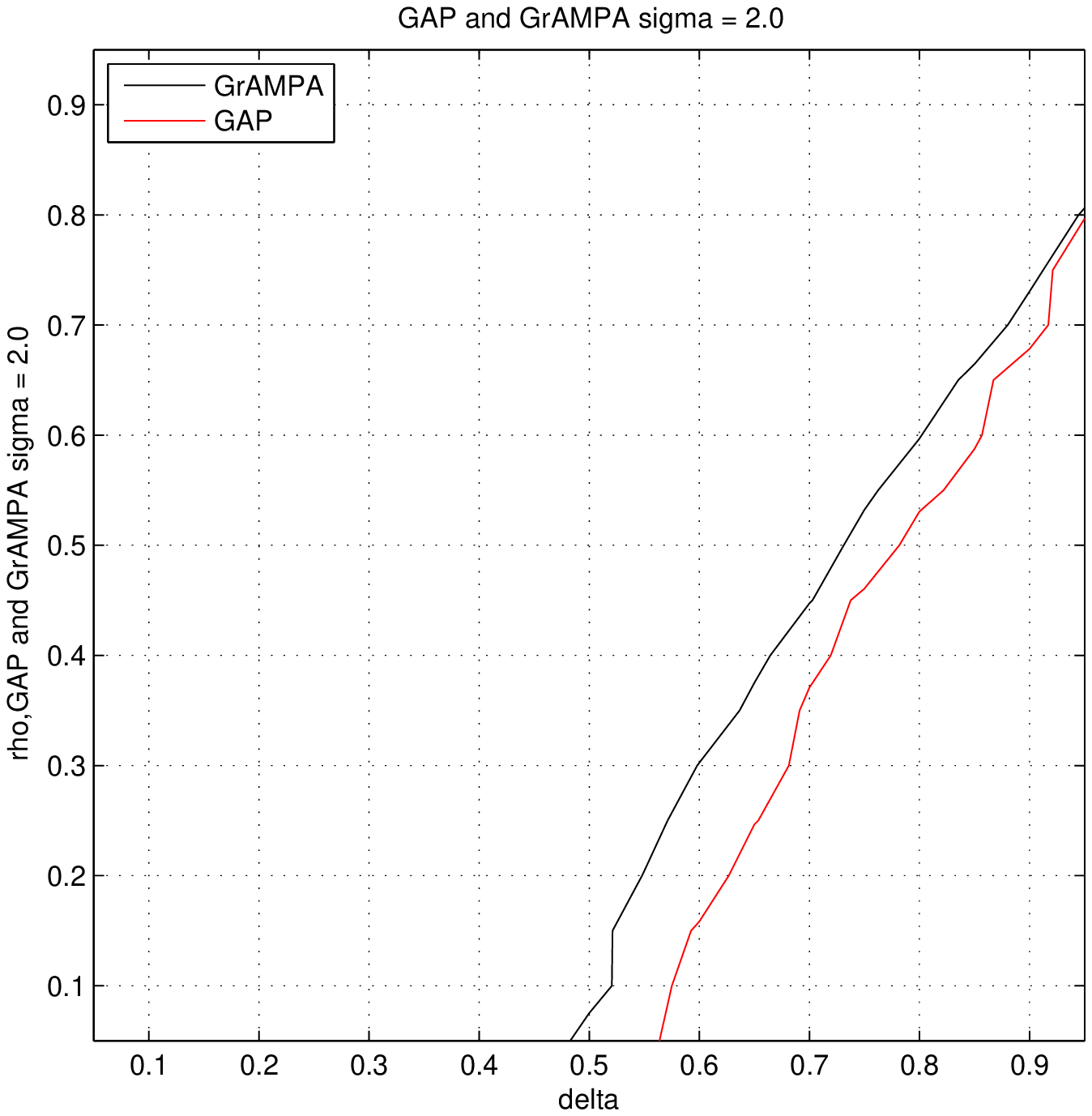}\\
   \end{tabular}
   \caption{Phase transition curves for recovery of $L$-cosparse $N$-length signals from $M$ noiseless measurements under i.i.d Gaussian $\vec{\Phi}$ and an $N\times D$ random, almost-uniform, almost-tight frame $\vec{\Omega}\tran$.}
   \label{fig:grampa_gap_ptc}
   \vspace{-3mm}
 \end{center}
\end{figure}

\Figref{grampa_gap_ptc} shows the empirical phase-transition curves (PTCs) for GAP and GrAMPA versus sampling ratio $\delta=M/N$ and uncertainty ratio $\rho=(N-L)/M$. 
For points below the PTC, recovery was successful with high probability, while for points above the PTC, recovery was unsuccessful with high probability. 
Here, we defined ``success'' as $\NSNR \geq 10^6$. 
%
\Figref{grampa_gap_ptc} shows that  
the PTC of GrAMPA is uniformly better than that of GAP.
It also shows that, for both algorithms, the PTC approaches the feasibility boundary (i.e., $\rho\!=\!1$) as $M/N\!\rightarrow\! 1$ but that, as the analysis operator becomes more overcomplete (i.e., $D/N$ increases), the PTC progressively weakens.

\subsection{Compressive image recovery via sparsity averaging}

Next, we repeat an experiment from \cite{Carrillo:SPL:13}, where the $N=512\!\times\! 512$ Lena image $\vec{x}$ 
was recovered from $M$ noisy complex-valued measurements $\vec{y}\!=\!\vec{\Phi x}+\vec{w}$ 
at $\SNR\!=\!40$~dB.
The measurements were of the ``spread spectrum'' form: $\vec{\Phi}=\vec{M F C}$, where $\vec{C}$
was diagonal with random $\pm 1$ entries, 
$\vec{F}$ was an $N$-FFT, 
and $\vec{M}\in\{0,1\}^{M\times N}$ 
contained rows of $\vec{I}_N$ selected uniformly at random.
An overcomplete dictionary $\vec{\Psi}\in\Real^{N\times 8N}$ was constructed from a horizontal concatenation of the first $8$ Daubechies orthogonal DWT matrices, yielding the analysis operator 
$\vec{\Omega}\!=\!\vec{\Psi}\tran$.
The use of highly overcomplete concatenated dictionaries is dubbed ``sparsity averaging'' in \cite{Carrillo:SPL:13}.


\putFrag{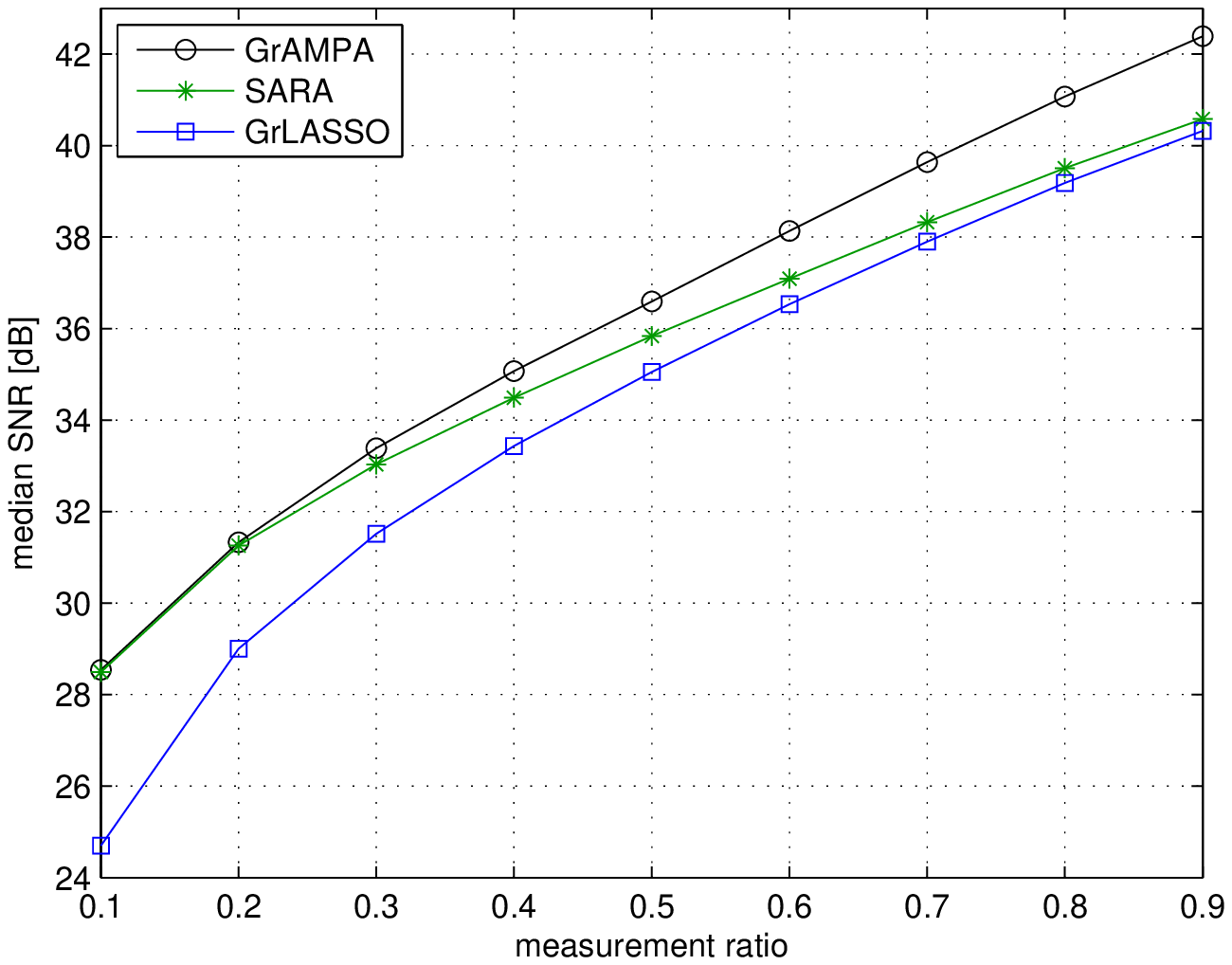}
{Recovery of the $512\times 512$ Lena image under a measurement SNR of $40$~dB, spread-spectrum $\vec{\Phi}$, and Db1-8 concatenated $\vec{\Omega}$.}
{\figsize}
{\psfrag{median SNR [dB]}[b][b][0.8]{\sf median recovery NSNR [dB]} 
 \psfrag{measurement ratio}[t][t][0.8]{\sf sampling ratio $M/N$}
}

\Figref{sara_vs_grampa_Lena512} shows median $\NSNR$ (over $30$ Monte-Carlo trials) versus sampling ratio $M/N$ for GrAMPA with SNIPE denoising; 
for ``SARA'' from \cite{Carrillo:SPL:13}, which employs iteratively-reweighted-$\ell_1$ \cite{Candes:JFA:08}; 
and for 
GrLASSO implemented via the ``SOPT'' Matlab code that accompanies \cite{Carrillo:SPL:13},
which employs Douglas-Rachford splitting \cite{Combettes:JSTSP:07}.
All algorithms enforced non-negativity in the estimate.
\Figref{sara_vs_grampa_Lena512} shows GrAMPA outperforming the other algorithms in $\NSNR$ at all sampling ratios $M/N$. 
%
Averaging over trials where all algorithms gave recovery $\NSNR\geq 30$~dB, the runtimes of GrAMPA, GrLASSO, and SARA were $220$, $255$, and $2687$ seconds, respectively.


\subsection{Shepp-Logan phantom recovery via 2D finite-differences}

Finally, we investigated the recovery of the $N=64\!\times\! 64$ Shepp-Logan Phantom image 
from 2D Fourier radial-line measurements $\vec{y}=\vec{\Phi x}+\vec{w}$ at $\SNR=80$~dB, using an analysis operator $\vec{\Omega}$ composed of horizontal, vertical, diagonal, and anti-diagonal 2D finite differences, as described in the noise-tolerant GAP paper \cite{Nam:CAMSAP:11}.

\Figref{radial_shepp_snr_vs_lines_64} plots median recovery $\NSNR$ (over 11 Monte-Carlo trials) versus number of radial lines for 
GrAMPA with SNIPE denoising,
GAPn \cite{Nam:CAMSAP:11}, 
the ``RW-TV'' approach from \cite{Carrillo:SPL:13}, which employs iteratively-weighted-$\ell_1$ \cite{Candes:JFA:08}, 
and GrLASSO, implemented using the Douglas-Rachford based ``SOPT'' Matlab code from \cite{Carrillo:SPL:13}.
The figure shows that GrAMPA achieved the best phase transition and also the best $\NSNR$ (for all numbers of radial lines above $6$).
Averaging over trials where all algorithms gave recovery $\NSNR\geq 30$~dB, the runtimes of GrAMPA, GrLASSO, RW-TV, and GAP were $0.28$, $1.8$, $9.7$, and $30.1$ seconds, respectively.

\putFrag{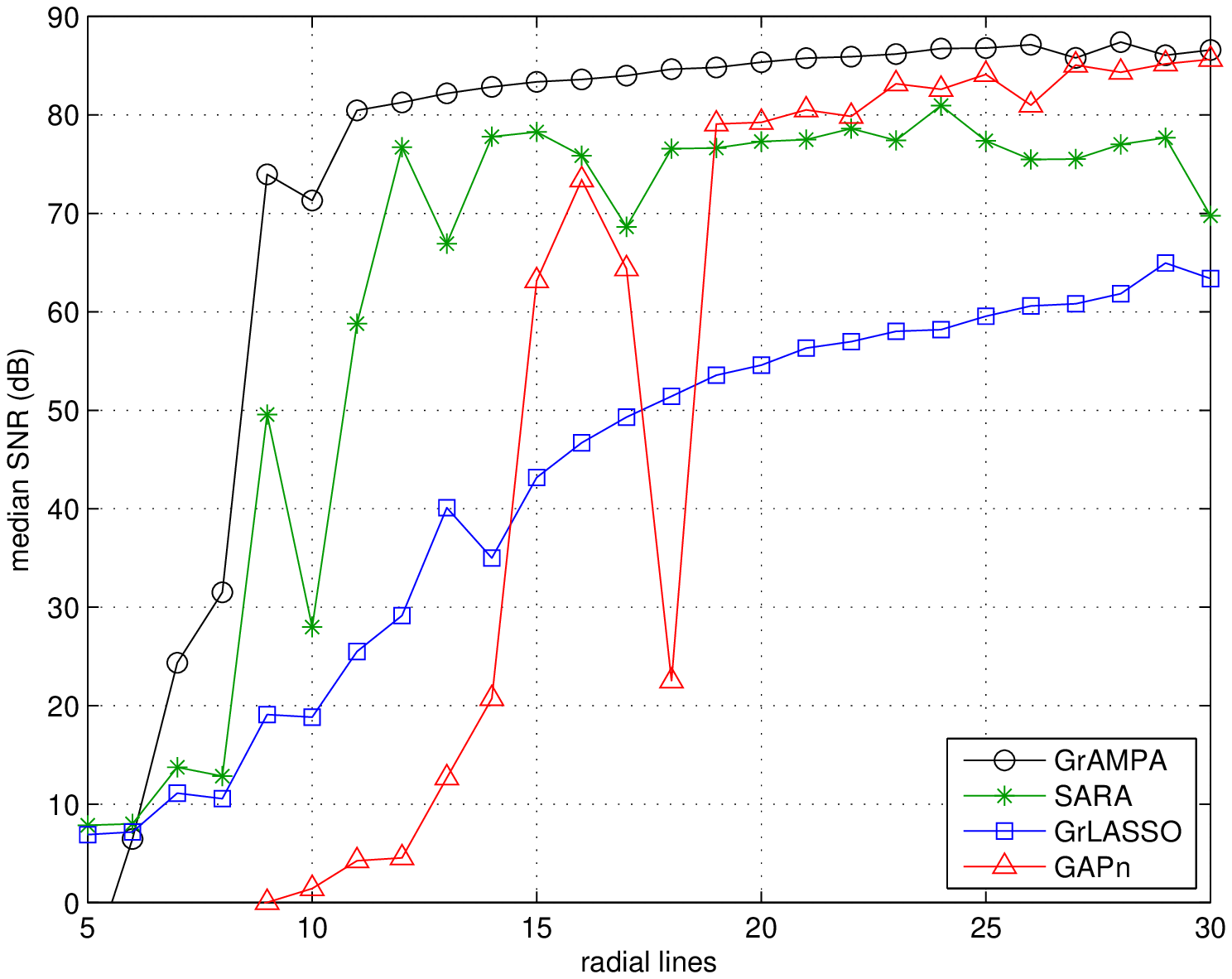}
{Recovery of the $64\times64$ Shepp-Logan phantom from 2D FFT $\vec{\Phi}$ and 2D horizontal, vertical, and diagonal finite-difference $\vec{\Omega}$.}
{\figsize}
{\psfrag{SARA}[l][l][0.58]{\sf \hspace{-0.2mm}RW-TV} 
 \psfrag{median SNR (dB)}[b][b][0.8]{\sf median recovery NSNR [dB]} 
 \psfrag{radial lines}[t][t][0.8]{\sf \# radial lines}}
{}

\section{Conclusions}

In this work, we proposed the ``Generalized AMP for Analysis CS'' (GrAMPA) algorithm, a new AMP-based approach to analysis CS that can be used with a wide range of loss functions, regularization terms, and analysis operators.
In addition, we proposed the ``Sparse Non-informative Parameter Estimator'' (SNIPE), an $\ell_0$-like soft thresholder that corresponds to the MMSE denoiser for a spike-and-slab distribution with an infinite-variance slab.
Numerical experiments comparing GrAMPA with SNIPE to several other recently proposed analysis-CS algorithms show improved recovery performance and excellent runtime.
Online tuning of the SNIPE parameter $\omega$ will be considered in future work.


\clearpage
\bibliographystyle{IEEEbib}
\bibliography{macros_abbrev,books,misc,sparse,machine,phase}
\end{document}